\documentclass[%
footinbib,
aip,
floatfix,
amsmath,amssymb,
reprint,%
]{revtex4-2}

\usepackage[utf8]{inputenc}
\usepackage[T1]{fontenc}
\usepackage{mathptmx}
\usepackage{etoolbox}
\usepackage{textcomp}

\usepackage[x11names]{xcolor}
\usepackage{graphicx}
\usepackage[normalem]{ulem}
\usepackage{soul}
\usepackage{dcolumn}
\usepackage{bm}
\definecolor{darkgreen}{RGB}{20,150,40}

\def\CORR#1{{#1}}

\usepackage{comment}
\usepackage{amsfonts}
\usepackage{amsmath}
\usepackage{stackengine}
\usepackage{wrapfig}
\usepackage{vwcol}
\usepackage{appendix}

\makeatletter
\def\@email#1#2{
 \endgroup
 \patchcmd{\titleblock@produce}
  {\frontmatter@RRAPformat}
  {\frontmatter@RRAPformat{\produce@RRAP{*#1\href{mailto:#2}{#2}}}\frontmatter@RRAPformat}
  {}{}
}
\makeatother

\begin{document}

\preprint{AIP/123-QED}

\title{A birdcage resonant antenna for helicon wave generation in TORPEX}

\author{Simon Vincent}
\author{Philippe Guittienne}
\author{Patrick Quigley}
\author{Cyrille Sepulchre}
\author{Rémy Jacquier}
\author{Robert Bertizzolo}
\author{Marcelo Baquero-Ruiz}
\author{Alan A. Howling}
\author{Ivo Furno}

\affiliation{École Polytechnique Fédérale de Lausanne (EPFL), Swiss Plasma Center (SPC), CH-1015 Lausanne, Switzerland}

\date{\today}
             
\begin{abstract}
A birdcage resonant helicon antenna is designed, mounted and tested in the toroidal device TORPEX. The birdcage resonant antenna is an alternative to the usual Boswell or half-helical antenna designs commonly used for $\sim 10$~cm diameter helicon sources in low temperature plasma devices. The main advantage of the birdcage antenna lies in its resonant nature, which makes it easily operational even at large scales, an appealing feature for the TORPEX device \CORR{whose} poloidal cross section is 40~cm in diameter. With this antenna helicon waves are shown to be launched and sustained \CORR{throughout} the whole torus of TORPEX. The helicon waves can be launched at low power on a pre-existing magnetron-generated plasma with little effect on the density profiles. The birdcage antenna can also be used alone to produce plasma, which removes the constraint of a \CORR{narrow range of applied magnetic fields} required by the magnetron, opening the way to a new range of studies on TORPEX with the external magnetic field as a control parameter.
\end{abstract}

\pacs{Valid PACS appear here}

\maketitle

\section{Introduction}
\label{sec::intro}

Helicon waves are bounded whistler waves, belonging to the right hand polarized part of electromagnetic waves that can propagate \CORR{in magnetized plasmas}, at frequencies between the ion and electron cyclotron frequencies $\omega_{ci} \ll \omega \ll \omega_{ce}$.
Helicon waves have attracted a lot of interest from the 1970s in the context of low-temperature plasmas, as they were found to have a surprisingly high ionization efficiency~\cite{Boswell_1970}: for sufficient input power the energy transfer from the wave to the plasma abruptly increases, reaching the so-called "helicon mode"~\cite{Shinohara_2018}.
Since then, a lot of experimental and theoretical studies have been carried out to better understand helicon properties and energy transfer capabilities. The higher ionization efficiency of helicon sources over the conventional ICP (inductively coupled plasma) sources, in configurations comprising an external magnetic field, is still a subject of active research. In the meantime, helicon sources have been developed across the globe and are now \CORR{routinely} used for plasma generation in laboratory devices~\cite{Thakur_2015, Scime_2007, Bohlin_2014, Furno_2017, Brochard_2023}, as well as in industry~\cite{Shinohara_2018}. 
Low-temperature plasma helicon sources can also be used \CORR{for} wall conditioning in tokamaks~\cite{Huang_2020}.
In the context of fusion, helicon waves have also been considered as a promising candidate for current drive in tokamaks~\cite{Vdovin_2013, Prater_2014}.
A number of numerical simulations of helicon wave propagation and absorption efficiency were performed in fusion relevant conditions~\cite{Lau_2018, Li_2020, Wu_2023} to predict the performances of helicon current drive as well as for antenna design purposes.
A recently mounted planar helicon antenna is currently being tested on the DIII-D tokamak~\cite{VanCompernolle_2021}, another one has been mounted on the tokamak KSTAR~\cite{Wi_2023}. 
\CORR{Experimental investigations of helicon wave physics in a toroidal geometry can nevertheless greatly benefit from fundamental studies in low temperature basic plasma devices, which can be performed in very well diagnosed systems at a lower cost.}

Helicon sources \CORR{have been} installed in a few low temperature plasma toroidal devices~\cite{Tripathi_2001, Grulke_2001, Sakawa_2004} and stellarator~\cite{Zhang_1995}, but few studies focused on the helicon waves themselves. 
Refs.~\cite{Zhang_1995, Paul_2005, Paul_2010} provide an interesting set of helicon wave amplitude measurements in toroidal configurations, \CORR{in devices with} major radii $\lesssim 30$~cm.
Nonetheless, experimental studies of helicon waves have been mostly restricted to cylindrical devices. To the knowledge of the authors, little is known about the effect of a toroidal geometry on the fundamental properties of helicon waves.
\CORR{To bridge this gap, we have designed and installed} a helicon source in TORPEX~\cite{Fasoli_2019}, \CORR{a toroidal device of major radius 1~m and minor radius 20~cm}. 

\CORR{In all previous studies, TORPEX relied on microwaves, generated by a magnetron, to create and sustain a plasma.} To this end the magnetron frequency has to match the electron cyclotron frequency $\omega_{ce}$ or lower hybrid frequency $(\omega_{ce}^2 + \omega_{pe}^2)^{1/2}$ (see Ref.~\cite{Podesta_2005}), which restricted the studies in TORPEX to values \CORR{of the total magnetic field} $B \in [650 ; 850]$~G. This further motivated the helicon antenna, \CORR{whose operation enables experiments in a very wide range of values of B,} 
hence opening the possibility for a whole new range of studies in TORPEX with $B$ as a control parameter.

This article presents the installation and first results of the newly mounted helicon source in TORPEX. The device TORPEX is presented in Sec.~\ref{sec::setup} along with the diagnostics used in the present work. \CORR{The antenna design that is chosen is that of a birdcage resonant helicon antenna, or bircage antenna (BCA).} The BCA is then introduced in \CORR{Sec.~\ref{sec::birdcage}, together with a thorough explanation of the advantages provided by this design.}
In Sec.~\ref{sec::propag_helicon}, the successful propagation of helicon waves launched by the BCA in TORPEX is demonstrated. Section~\ref{sec::plasma_birdcage} finally presents a \CORR{series} of measurements establishing the BCA ability to sustain plasma in various conditions.  
Conclusions are \CORR{finally given} in Sec.~\ref{sec::conclusion}.

\section{TORPEX device and diagnostics}

\label{sec::setup}

\begin{figure}
    \centering
    \includegraphics[width = 0.98\columnwidth, trim={0in 0in 0in 0in},clip]{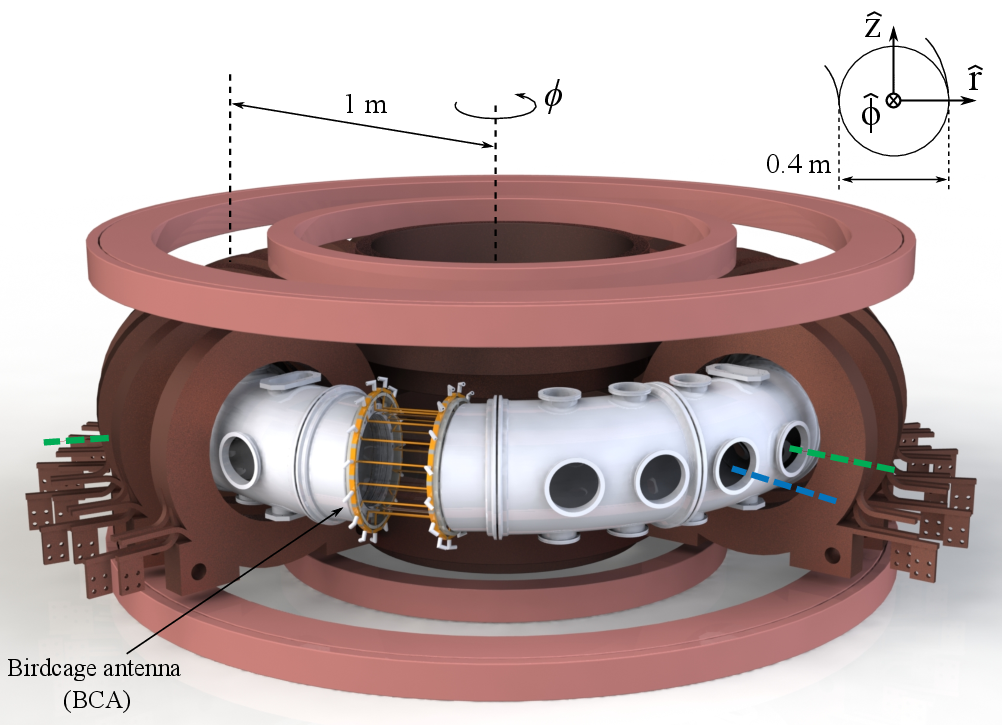}
    \caption{TORPEX device, with the BCA.
    \CORR{The Langmuir probe (blue dashed line) is inserted at the toroidal location $\phi \approx + 80 \, ^{\circ}$, and the B-dot probes (green dashed lines) at the locations $\phi \approx -80 \, ^{\circ}$ and $\phi \approx + 100 \, ^{\circ}$,} on the left \CORR{or} right sides of the antenna respectively.
    \CORR{Note that toroidal coils on TORPEX surround the whole chamber: a set of these coils are not shown in the figure for a better visualisation of the chamber.} }
    \label{torpex_sketch}
\end{figure}

TORPEX is a toroidal device of major and minor radii $1$~m and $20$~cm respectively (Fig.~\ref{torpex_sketch}). We use here a simple toroidal coordinate system $(\hat{r}, \hat{\phi}, \hat{z})$ \CORR{as shown in Fig.~\ref{torpex_sketch}}, \CORR{with the origin $(0, 0, 0)$ taken at the center of the BCA.} The stainless steel chamber of TORPEX is surrounded by a set of 28 \CORR{toroidal} coils and 5 pairs of \CORR{poloidal} coils, enabling a wide variety of magnetic configurations, with values of the total magnetic field \CORR{potentially} up to $1000$~G. 
The magnetic configuration chosen for this study is the so-called Simple Magnetized Torus (SMT)~\cite{Muller_2004, Fasoli_2006} that consists of a dominant toroidal field with a smaller vertical field, of values $B_{tor} \approx 780$~G and $B_{ver} \approx 20$~G respectively (averaged over the poloidal cross section).
Hydrogen or argon plasmas \CORR{can be} generated by a 2.45~GHz magnetron, \CORR{upon injection of the corresponding gases into the vessel to a pressure of} $\sim 10^{-4}$~mbar up to $\sim 10^{-3}$~mbar.
\CORR{The main vessel of TORPEX} is composed of 8 toroidal \CORR{sectors}. One of these \CORR{sectors} was replaced by two flanges and a borosilicate tube surrounded by a BCA, \CORR{whose} technical details and \CORR{capabilities are} the subject of Sec.~\ref{sec::birdcage}. 


The plasma density $n_e$ and floating potential $V_f$ are measured with a Langmuir probe, scanning radial positions at $z=0$~cm and at a toroidal angle $\phi \approx +80 \, ^{\circ}$ away from the antenna center (see blue dashed line in Fig.~\ref{torpex_sketch}).
The density is \CORR{obtained} from measurements of the ion saturation current $I_{i,sat}$, that is collected with a probe tip \CORR{biased at} $V_{bias} = -60$~V. We then use $n_e = |I_{i,sat}|/ (\alpha_0 e A \sqrt{e T_e/m_i} \beta$), with $\alpha_0=0.5$ a coefficient taking into account the effect of the pre-sheath~\cite{book_Lieberman, Furno_2014}, $e$ the electron charge, $A$ the probe tip surface, $T_e$ the electron temperature assumed here constant at 4~eV (measurements in similar conditions show $T_e \sim 2-5$~eV), $m_i$ the ion mass, and $\beta = (1 - \gamma (V_{bias} - V_f))$ a coefficient accounting for the sheath expansion. 
Based on results from previous studies~\cite{Theiler_2011}, we use here $\gamma \sim 0.05$.

A magnetic probe (dubbed B-dot probe) is used to measure the magnetic field fluctuations along all axes $(\hat{r}, \hat{\phi}, \hat{z})$ at any \CORR{selected location}. 
The B-dot probe consists \CORR{of} three orthogonal 5~mm $\times$ 5~mm square coils, each made of 10 loops of 0.2~mm coated copper wire. The probe's head is mounted on a ceramic tube and inserted in a closed-end glass tube \CORR{to provide protection from direct contact with} the plasma. The coil ends are connected from the probe's head to hybrid combiners, from which the B-dot measurement is acquired. Note that the connections from the 6 ends of the probe's head to the hybrid combiners are done with individual coaxial cables, which protects the measurement from capacitive or inductive pick-up along this transmission line.

\begin{figure}
    \centering
    \includegraphics[width = 0.98\columnwidth, trim={0in 0in 0in 0in},clip]{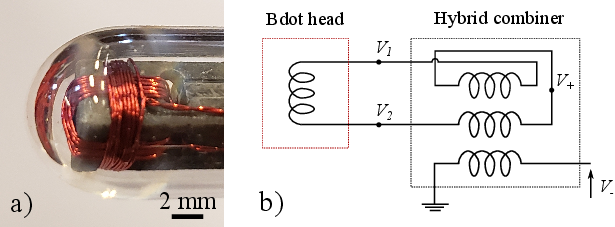}
    \caption{a) B-dot probe head. b) Electrical sketch of one of the coils of the B-dot, with a hybrid combiner measuring $V^-~=~(V_1 - V_2)/2$, leaving the B-dot head electrically floating in the plasma.}
    \label{Bdot_sketch}
\end{figure}

A picture of the probe's head with the coils is shown in Fig.~\ref{Bdot_sketch} (a) and the \CORR{circuit diagram} corresponding to each coil is drawn in Fig.~\ref{Bdot_sketch} (b). 
As shown in Fig.~\ref{Bdot_sketch} (b) each one of the three B-dot coils (with given potentials $V_1$ and $V_2$ at its ends) is connected to a hybrid combiner, \CORR{composed of three coils winded together, hence providing the measurement of}  $V^- = (V_1 - V_2)/2$. This way the induced current in the coil is directly \CORR{obtained}, removing the potential capacitive pick-up from the plasma (that would affect in the same way $V_1$ and $V_2$). In addition using hybrid combiners enables the B-dot coils to be \CORR{electrically} floating and unperturbed by the acquisition system.

B-dot measurements were performed along $r$ and at $z=0$~cm, at the toroidal angles $\phi \approx -80 \, ^{\circ}$ and $\phi \approx +100 \, ^{\circ}$ away from the antenna center, as shown in green dashed lines in Fig.~\ref{torpex_sketch}. More details on the B-dot probe calibration are provided in appendix~\ref{appendix::Bdot}. As will be presented in section~\ref{sec::propag_helicon}, B-dot measurements reveal the presence of helicon waves generated by the BCA.

\section{The Birdcage resonant helicon antenna}
\label{sec::birdcage}

Since the discovery \CORR{of the high ionization efficiency of helicon sources, numerous designs have been developed.} Among these are the well known Boswell antenna~\cite{Boswell_1970}, or the half helical antenna~\cite{Miljak_1998} that is now the most commonly used one in linear devices, where the antenna diameters are typically $d \lesssim 10$~cm. All these designs essentially consist \CORR{of} wire loops, \CORR{which are inductive elements.}
\CORR{Even when coupled to the generated plasma, whose impedance is partly resistive due to collisions, these helicon plasma sources have an impedance that is mostly inductive. When fed by a RF power supply, the antenna input is therefore prone to high voltages.} \CORR{This can lead to} arcing problems, making plasma generation difficult to achieve when the size of the antenna is increased. 
It is however worth mentioning here the work of Ref.~\cite{Shesterikov_2019} where a very large half helical antenna of 44~cm diameter could be used for plasma generation, and even reach helicon mode coupling as reported by the authors.
The RF engineering difficulties that had to be overcome to get these results are not detailed in the article.

In the case of TORPEX, \CORR{we chose an alternative helicon antenna design, the BCA, which advantages} are explained in Sec.~\ref{subsec::birdcage_advantages}. The choice of the specific BCA design to be mounted on TORPEX is then motivated by COMSOL simulations in Subsec.~\ref{subsec::num_simu}. The final design and \CORR{installation} of the BCA on TORPEX is presented in Subsec.~\ref{subsec::birdcage_design}.

\subsection{Description and advantages of the BCA}
\label{subsec::birdcage_advantages}

\begin{figure}
    \centering
    \includegraphics[width = 0.98\columnwidth, trim={0in 0in 0in 0in},clip]{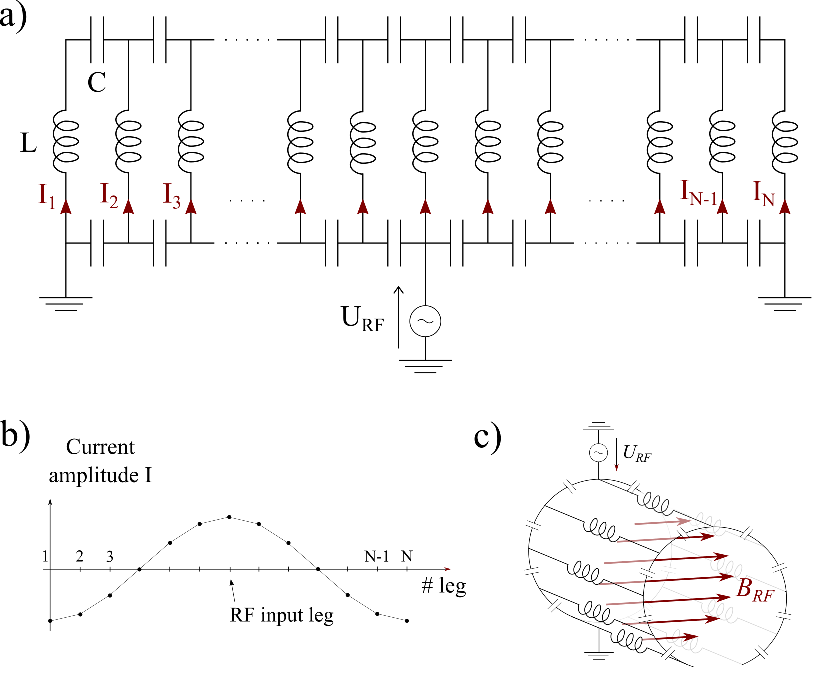}
    \caption{a) Electrical circuit of the birdcage antenna resonant network. b) Sinusoidal distribution of currents amplitudes in the legs of a BCA, \CORR{when excited at the resonance frequency corresponding to modes $|m|=1$}. c) Drawing of the resonant network \CORR{folded around} to form a birdcage antenna, with the transverse magnetic field induced at the resonance corresponding to $|m|=1$.}
    \label{electrical_sketch_birdcage}
\end{figure}

The key feature of a BCA is that it is a resonant device. This \CORR{type of} antenna consists of an assembly of legs of inductance L connected in parallel by \CORR{capacitors of capacitance} C, as shown in Fig.~\ref{electrical_sketch_birdcage} (a). This \CORR{circuit} has resonance frequencies $\omega_{res}^p \sim 1/\sqrt{2 \, L \, C \sin^2(\frac{p \, \pi}{2 \,  N})}$ (see Ref.~\cite{Guittienne_2014}) with N the number of inductive legs and $p \in [1; N-1]$.

\CORR{At resonance, the input impedance $Z$ of such a circuit becomes mostly real, which is a major difference compared to the classical helicon antennae mentioned before. 
When the excitation frequency matches a resonance frequency, the real part of the impedance $\text{Re}(Z)$ typically reaches} a value \CORR{of the order of} a few hundred Ohm \CORR{(when coupled to plasma  $\text{Re}(Z)$ is even further reduced to few tens of Ohm)}, and the imaginary part $\text{Im}(Z)$ is \CORR{small (a few Ohm) to negligible}. \CORR{Therefore} the phase shift between the input voltage $U_{in}$ and current $I_{in}$ is close to zero, and the input power that is \CORR{dissipated} by the antenna does not require high voltages nor high currents. Such an antenna is therefore scalable to larger dimensions than ICP or conventional helicon sources, \CORR{while} avoiding arcing problems. \CORR{As an example, with an ICP made of a 4 turns solenoid with diameter 13~cm, and powered by a 1~kW, we can expect $U_{in} \approx 9500 $~V and $I_{in} \approx 30$~A. With a BCA of diameter 13~cm and same input power, we only have $U_{in} \approx 140 $~V and $I_{in} \approx 7$~A (see Ref.~\cite{Guittienne_2024})}.

Being bound to operate at resonance can be seen as a significant inconvenience of this antenna design. In reality this aspect is not too restricting, because the BCA operation is in fact quite robust: even if the power supply frequency does not perfectly match the targeted resonance, we still have $\text{Re}(Z) \sim \text{Im}(Z)$. This is enough to substantially reduce the voltage levels compared to a situation where $\text{Re}(Z) \ll \text{Im}(Z)$, and hence scale up the antenna. 

To make a \CORR{BCA}, the \CORR{planar circuit} sketched in Fig.~\ref{electrical_sketch_birdcage} (a) is \CORR{rolled} in a cylindrical shape as shown in Fig.~\ref{electrical_sketch_birdcage} (c). The connection to the ground is made at the location \CORR{opposite} to the RF input (\CORR{details on the reason for this choice can be found in Ref.~\cite{Guittienne_2014}}). On the other side of this ground connection, along the cylinder axis, the circuit is left open.
\CORR{Selecting} the resonance frequency corresponding to $p = 2$ \CORR{results in a} current distribution in the legs \CORR{which} follows a sinusoidal pattern, \CORR{corresponding to an azimuthal mode $m=1$}. Note that in the following, $m$ refers to the poloidal wavenumber of a helicon wave of the form $e^{i (\omega t - m \theta - k_{\phi}\phi )}$, with $\theta = \sqrt{r^2 + z^2}$ the poloidal angle and $k_{\phi}$ the toroidal wavenumber. Arranged in a cylindrical shape, the legs then produce an homogeneous transverse magnetic field, that is able to excite helicon waves (see Fig.~\ref{electrical_sketch_birdcage} (b) and (c)).

An additional advantage of the BCA is that such a sinusoidal pattern is well suited for the exclusive excitation of the helicon modes $|m| = 1$. Indeed, a half-helical antenna having only 2 legs (whose current amplitudes would be the maxima in Fig.~\ref{electrical_sketch_birdcage} (b)) is able to excite all modes $m = 2q$, with $q\in Z$. \CORR{Thus,} the power this antenna transfers to the plasma is likely to be scattered among various modes $m$. \CORR{In contrast to this, the BCA benefits from a more targeted coupling to the modes $|m| = 1$, and avoids this spread in energy transfer, by imposing with N points \CORR{an azimuthal} wavelength corresponding to modes $|m| = 1$.}
This is an appealing feature of the BCA excitation, since the helicon mode $m = +1$ is known to be the most efficient for plasma ionization~\cite{Shinohara_2018}.

The geometry of the BCA, with its legs wrapped around a cylinder, ensures moreover an important surface of interaction between the antenna currents and the plasma. 
The induction mode of a BCA is therefore easily achieved, even at low pressures compared to classical helicon antennae. 

\CORR{Another advantage is that} the impedance characteristics of the BCA (\CORR{$\text{Re}(Z) \sim 100$~$\Omega$} and $\text{Im}(Z) \lesssim \text{Re}(Z)$) make it easier to be matched to 50~$\Omega$. This leads to lower levels of currents \CORR{and} less power dissipation in the matching-box \CORR{(which is the intermediate element composed of variable capacitors placed between the RF power supply and the antenna, with which the matching is achieved).} This also means that the coaxial cable \CORR{connecting} the matching-box to the BCA 
has to withstand a lower value of reflected power
and is therefore less subjected to power dissipation and heating. This results in a better antenna efficiency, \CORR{and makes in addition} the antenna system easier to build and \CORR{more practical to operate}.

The main advantages of a BCA over the classical helicon antenna designs can be summarized as follows:
\begin{itemize}
    \item Low input voltages and currents resulting from the resonant nature of the antenna, making it easily scalable to large dimensions.
    \item Sinusoidal distribution of currents in the legs, providing a better targeted energy deposition to helicon modes $|m| = 1$.
    \item Efficient inductive coupling and easy plasma ignition, thanks to the large surface covering of the high RF energy stored in the legs.
    \item \CORR{Low} power dissipation in the matching-box and input cable.
\end{itemize}

\CORR{We note that these advantages come, of course, at the expense of a more complex design.}
A more detailed comparison between a BCA and a half-helical antenna in a simple linear device will be the subject of a forthcoming publication. For an in-depth description of the physics and engineering of the BCA, we refer \CORR{the reader} to Ref.~\cite{Guittienne_2024}.

\subsection{Numerical simulations}
\label{subsec::num_simu}

The \CORR{installation} of the BCA in TORPEX is done \CORR{with} a \CORR{cylindrical} glass tube of diameter 30~cm, connected to the toroidal vessel via two flanges \CORR{that were designed to adapt the straight cylinder to the toroidal geometry of TORPEX.}
The \CORR{geometry} of the BCA is constrained on the inner side by the glass tube, and on the outer side by the toroidal coils of inner diameter 52~cm.
\CORR{With these spatial constraints}, the value of the capacitances $C$ of the antenna, as well as the length and shape of the legs (defining the value of $L$), have to be chosen for the $|m|=1$ resonance to match the 13.56~MHz frequency of our power supply. \CORR{We note} in addition that the resonance frequency is expected to shift upwards in the presence of the plasma. \CORR{To choose the best antenna shape and components, taking into account the spatial constraints and this frequency shift,} numerical simulations with various BCA designs were performed with COMSOL\textregistered~\cite{COMSOL}.

\begin{figure}
    \centering
    \includegraphics[width = 0.98\columnwidth, trim={0in 0in 0in 0in},clip]{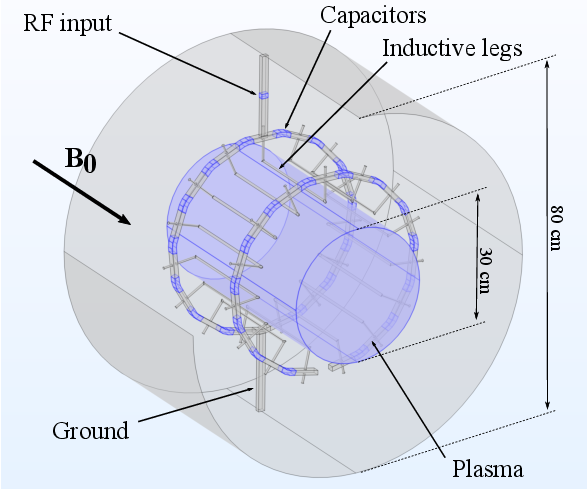}
    \caption{Geometry of the birdcage antenna and spatial domain of the COMSOL numerical simulations.}
    \label{comsol_sketch}
\end{figure}

The spatial domain for the simulations is a cylinder of diameter 80~cm and length 50~cm, \CORR{that includes} a plasma domain consisting of a cylinder of diameter 30~cm. These domains are shown in Fig.~\ref{comsol_sketch}, \CORR{together with} the BCA design that was finally chosen, and that will be discussed \CORR{in detail in Sec.~\ref{subsec::birdcage_design}}. The BCA is placed around the plasma domain, with the legs at 1.5~cm from the plasma (hence from the inner tube boundary).
The capacitors are treated as lumped elements, while the boundaries of the legs and the outer limits of the vacuum domain are perfectly conducting surfaces such that $\vec{n} \times \vec{E} = \vec{0}$, \CORR{where $\vec{n}$ is the unity vector perpendicular to the surface and $\vec{E}$ the electric field at the surface}. An external magnetic field $B_0$ is imposed along the \CORR{axial} direction of the cylindrical domain as shown in Fig.~\ref{comsol_sketch}. The plasma is treated as a dielectric, \CORR{and the electromagnetic field is modeled with}
$$\vec{\nabla} \times (\vec{\nabla} \times \vec{E})
    - k_0^2  \left(\boldsymbol{I} -  \frac{i}{\omega \varepsilon_0}\boldsymbol{\sigma} \right) \cdot \vec{E} = \vec{0} ,$$
with a conductivity $\boldsymbol{\sigma} = \boldsymbol{0}$ in vacuum, and in the plasma:
\begin{align*}
    \boldsymbol{\sigma}
    =  \alpha
    \Bigg( \Bigg.\begin{matrix}
    \gamma^2 & \gamma \omega_{ce} & 0 \\
    - \gamma \omega_{ce} & \gamma^2 & 0 \\
    0 & 0 & \gamma^2 + \omega_{ce}^2
    \end{matrix}\Bigg. \Bigg)
    \ \ \ \ \textrm{and} \ \ \ \
    \begin{cases}
        \alpha & = \frac{\varepsilon_0 \omega_{p}^2}{\gamma (\gamma^2 + \omega_{ce}^2)} \\
        \gamma & = \nu + i \omega
    \end{cases}
\end{align*}

\noindent with $k_0 = c/\omega$, \CORR{$c$ the speed of light, $\varepsilon_0$ the vacuum permittivity,} 
$\omega_p = \sqrt{n_e e^2/\varepsilon_0 m_e}$ the plasma frequency, 
\CORR{$n_e$ the plasma density, $e$ the electron mass, $m_e$ the electron mass,}
$\nu$ the electron neutral collision frequency. Here the neutral pressure is set at \CORR{$10^{-4}$~mbar}, and we take $\nu = n_n \sigma_{en} v_{th,e}$, with $n_n$ the neutral pressure, $v_{th,e}$ the electron thermal velocity, and $\sigma_{en} = 10^{-19}$~m$^2$ the electron neutral elastic collision cross section of hydrogen, taken from the LXCat database~\cite{LXCat_2023, Pitchford_2017}, and evaluated at $T_e \sim 5$~eV.

The plasma density is assumed to be uniform with $n_e = 5 \times 10^{16}$~m$^{-3}$, that is of the order of the maximal density expected on TORPEX \CORR{at this pressure} and for a power up to 1~kW. The high density taken here, for TORPEX standards, as well as taking a uniform profile, is a way of assessing the most unfavorable scenario for the antenna being impacted by the presence of the plasma. This allows us to estimate an upper limit for the frequency shift due to the plasma.

\begin{figure}
    \centering
    \includegraphics[width = 0.98\columnwidth, trim={0in 0in 0in 0in},clip]{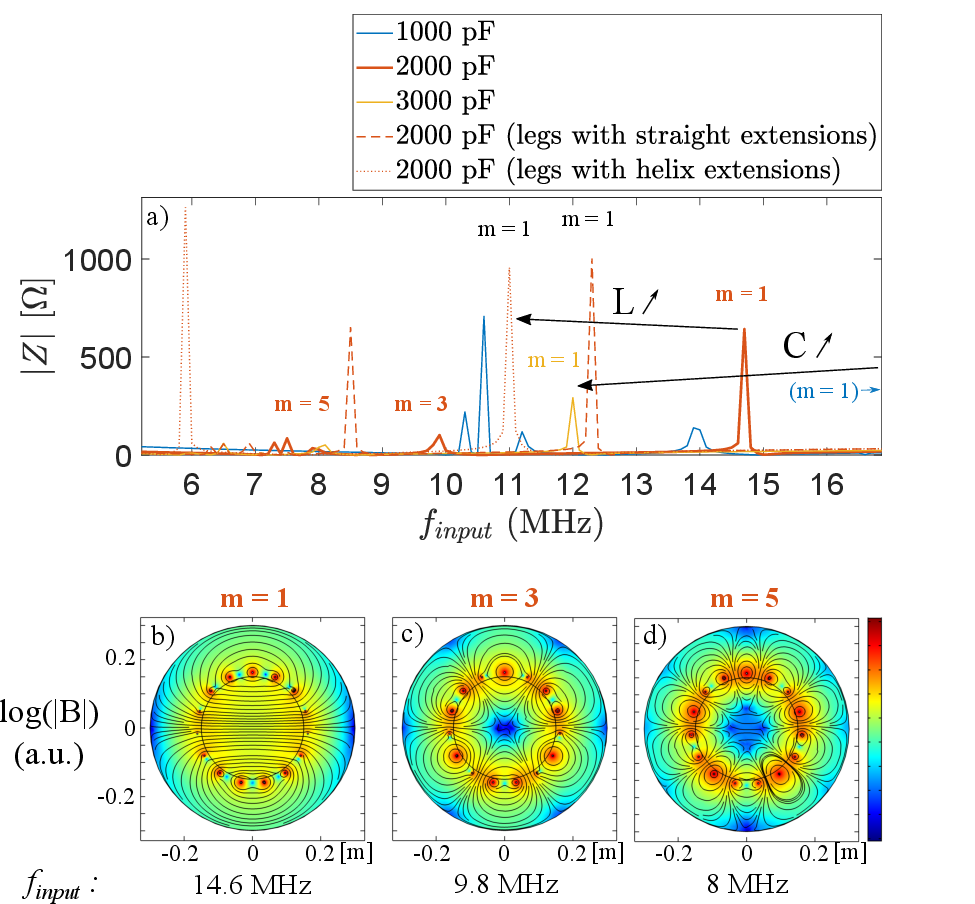}
    \caption{a) Input impedance spectrum of the birdcage resonant antenna, for values of the capacitance $C \in {1000, 2000, 3000}$~pF and three designs of the inductive legs: simple legs without extension, legs with straight extensions (see Fig.~\ref{comsol_sketch}), and legs with helix extensions. \CORR{b-c-d) Amplitude of the RF magnetic field simulated for an excitation frequency of the antenna corresponding respectively to modes $m=1$, $m=3$ and $m=5$.}}
    \label{Z_input_comsol}
\end{figure}

By performing series of simulations with a RF input frequency ranging from 5 to 17~MHz, the input impedance spectrum of the BCA can be obtained, showing various resonances corresponding to exciting modes $m=2n+1$ with $n\in \mathbf{N}$. Figure~\ref{Z_input_comsol} (a) shows the \CORR{modulus} of the BCA input impedance, for capacitances $C=2000$~pF and a simple design of the antenna with straight legs connecting the capacitors (red \CORR{solid line}). Figures~\ref{Z_input_comsol} (b), (c) and (d) show the amplitude of the RF magnetic field generated by the BCA at frequencies  corresponding to resonances $m=1$, $m=3$ and $m=5$, in a plane transverse to the BCA and at its center. As shown in Fig.~\ref{Z_input_comsol} (a), when the value of the capacitance is lowered to 1000~pF (blue line) or increased to 3000~pF (yellow line) the whole spectrum is \CORR{respectively} shifted upwards or downwards \CORR{in frequency}; \CORR{the resonances follow a trend} $\sim 1/\sqrt{LC}$. The value of the capacitance can hence be finely tuned to choose the resonance frequency.

However the COMSOL simulations performed with and without plasma, showed that the simple design of a BCA with straight legs led to a substantial upward shift of the resonance frequencies, shift that would in addition strongly depends on $B_0$ (with a shift of $\approx 2$~MHz with $B_0=0$~G, to $\approx 0.1$~MHz with $B_0=500$~G). This was considered to be impractical for studies on TORPEX that would imply varying $B_0$.
This effect of frequency shift in presence of the plasma can be reduced. To do so we limit the part of the BCA legs that are close and hence coupled to the plasma, with respect to the total length of the legs. This is done by adding extensions to the legs, away from the antenna center. With straight leg extension as shown in Fig.~\ref{comsol_sketch} for instance,  only $\approx 2/3$ of the legs are coupled to the plasma. This significantly reduces the resonance frequency shift to $\approx 0.3 - 0.7$~MHz for $B_0 \in [0;500]$~G. 
Note that of course by changing the leg total length, the value of $L$ is increased, leading to a change in the BCA spectrum as shown in Fig.~\ref{Z_input_comsol} (a) (dashed and dotted red curves). The leg design has therefore to be chosen first, and the value of the capacitance is then adjusted to choose the $m=1$ resonance frequency at $\approx 13.56$~MHz in the presence of plasma.

Overall various BCA designs have been explored, and a large number of simulations were performed, that will not be discussed in detail here. Although the frequency shift could be reduced even more, with longer leg extensions in the shape of helices for example (see red dotted curve in Fig.~\ref{Z_input_comsol} (a)), the final design is the one sketched in Fig.~\ref{comsol_sketch}, that combines a good reduction of the resonance frequency shift with a relative simplicity of the design. The BCA characteristics are detailed in subsection~\ref{subsec::birdcage_design}.

\subsection{Final design of the BCA and preliminary tests}
\label{subsec::birdcage_design}

\begin{figure}
    \centering
    \includegraphics[width = 0.98\columnwidth, trim={0in 0in 0in 0in},clip]{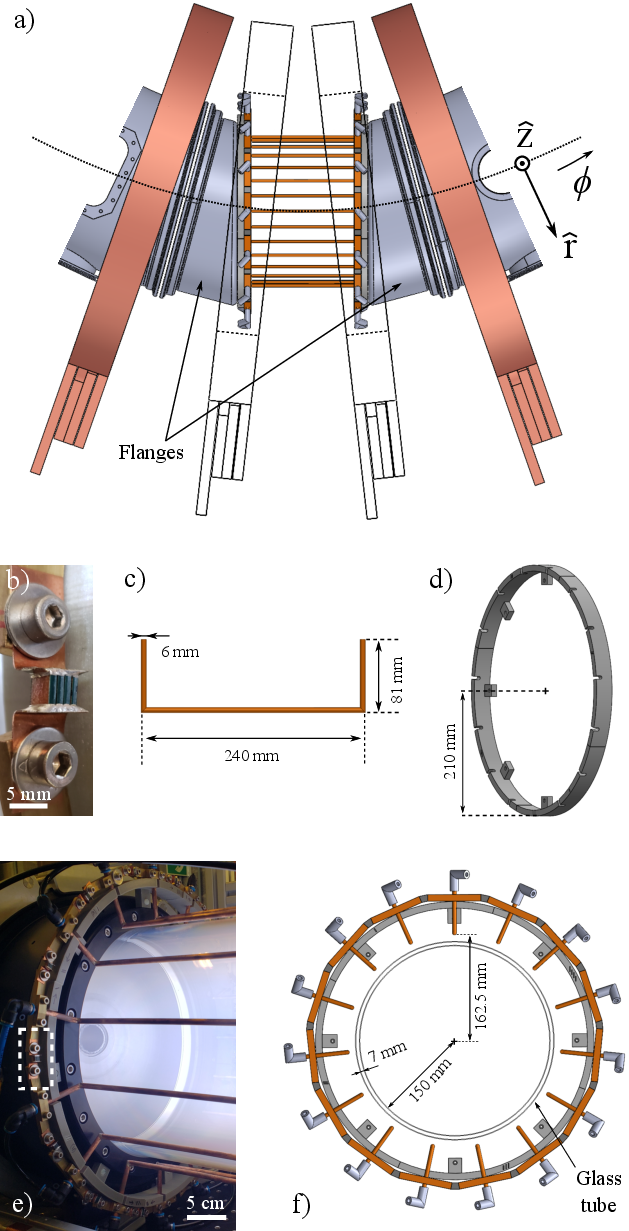}
    \caption{Elements of the BCA mounted on TORPEX. \CORR{See text for details.}}
    \label{birdcage_assembly}
\end{figure}

The final \CORR{design of the BCA and its installation} in TORPEX are presented in Fig.~\ref{birdcage_assembly}.
\CORR{The capacitances $C$ correspond to assemblies of four capacitors in parallel} (Fig.~\ref{birdcage_assembly} (b)), two of 430 pF and two of 470 pF adding up to a total capacitance of  1800 pF. Each capacitor can \CORR{withstand} up to 15~A rms without the need for active cooling: the assemblies can then support up to 60~A rms, corresponding to an input antenna power of up \CORR{to a few kW} in continuous mode. Without cooling of the capacitors, higher power can \CORR{still} be tested in pulse mode.
The legs are L-shaped (Fig.~\ref{birdcage_assembly} (c)) and connected at their ends to the capacitor assemblies via brass pieces (Fig.~\ref{birdcage_assembly} (e), (f)).

The BCA is supported by circular 3D printed pieces that are attached to the flanges on both sides (Fig.~\ref{birdcage_assembly} (c)).
Note that the \CORR{spacing} between the glass tube and the side flanges is \CORR{smaller} on the high field side than on the low field side (Fig.~\ref{birdcage_assembly} (a)), \CORR{which ensures} that the center of the BCA overlaps the main vessel toroidal central axis $\phi$. A matching-box of type T was built to connect the RF power supply to the BCA. This matching-box is composed of two sets of a $0-500$~pF variable capacitor and  $2.3$~$\mu$H \CORR{in-house made} coils connected in series, with a $100$~pF capacitor in between and connected to the ground (see Ref.~\cite{Guittienne_2024} for more details on type T matching-box).

\begin{figure}
    \centering
    \includegraphics[width = 0.98\columnwidth, trim={0in 0in 0in 0in},clip]{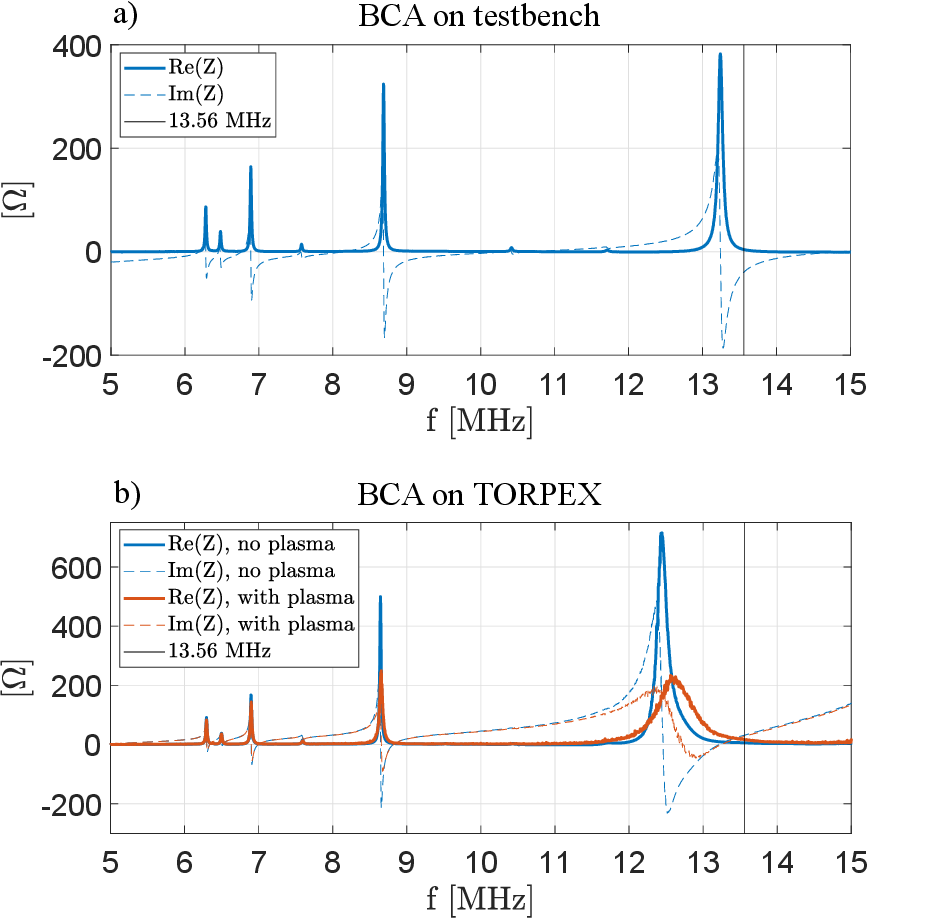}
    \caption{Input impedance of the BCA, measured on a testbench (a) and when mounted on TORPEX (b) with and without a background magnetron plasma.}
    \label{Zinput}
\end{figure}

The BCA was first mounted on a testbench composed of the glass tube and two flanges. \CORR{The measured input impedance (real and imaginary parts) of the antenna is shown} in Fig.~\ref{Zinput} (a). The clearly visible five \CORR{largest} peaks correspond to excitation modes $m=\{1, 3, 5, 7, 9\}$ of the RF transverse magnetic at \CORR{corresponding} frequencies $f=\{13.2, 8.7, 6.9, 6.5, 6.3\}$~MHz, with a highest resonance impedance for $m=1$ at \CORR{$|Z| \approx \Re(Z) \approx 380$~$\Omega$}.

\CORR{After initial tests, the antenna was} mounted on TORPEX, and the measured input impedance is shown in Fig.~\ref{Zinput} (b). The resonance frequency of mode $m=1$ is shifted down to $12.45$~MHz, with a corresponding resistive impedance peaking at 710~$\Omega$. This downward shift of the spectrum \CORR{is expected to be due to the close proximity to the magnetic field coils, the vacuum vessel, and other conducting materials.}
To \CORR{study} the impact of the plasma on the impedance, argon plasma is generated with the magnetron at power 600~W, and with an external field of \CORR{$B_{tor} \approx 780$~G}. As shown in the next subsection this plasma reaches \CORR{a} density of $\sim 10^{16}$~m$^{-3}$ and is localized near the high field side around $-12 \leq r \leq -2$. Figure~\ref{Zinput} (b) shows that in the presence of plasma the resonance frequency is slightly shifted upwards to 12.7~MHz. The peak of $\text{Re}(Z)$ is enlarged and its maximum value \CORR{reduced} to 200~$\Omega$.

With the strong impact of TORPEX environment \CORR{surrounding elements}, \CORR{and even with the upward frequency shift due to the presence of the plasma,} the resonance frequency is then $\approx 1$~MHz lower than the frequency \CORR{originally planned} of 13.56~MHz that is set by the RF power supply. The capacitor assemblies could be changed to correct this, but as will be seen in section~\ref{sec::propag_helicon} such a change is not necessary for the \CORR{correct operation} of the BCA.
At 13.56~MHz we have in the presence of plasma $\text{Re}(Z)\approx 20$~$\Omega$ and $\text{Im}(Z) \approx 30$~$\Omega$. In these conditions the matching of this input impedance to $50$~$\Omega$ is easily achieved, and draws low currents in the matching-box \CORR{and} in the cable connecting the matching-box to the BCA. The connection cable does not \CORR{require water cooling}, which simplifies the mounting of the BCA in TORPEX. Once the BCA mounted, its ability to launch and sustain helicon waves, as well as to generate plasma in TORPEX, is explored.

\section{Propagation of toroidal helicon waves}
\label{sec::propag_helicon}

First measurements of helicon waves excited by the BCA in TORPEX are now presented, in magnetron-generated hydrogen plasmas at a pressure of $\sim 10^{-4}$~mbar. The magnetic configuration is the SMT configuration introduced in Sec.~\ref{sec::setup}. The magnetron power is set at $P_{MAG} = 600$~W, and the BCA is fed with an RF power of $P_{BC} = 200$~W.

\begin{figure}
    \centering
    \includegraphics[width = 0.98\columnwidth, trim={0in 0in 0in 0in},clip]{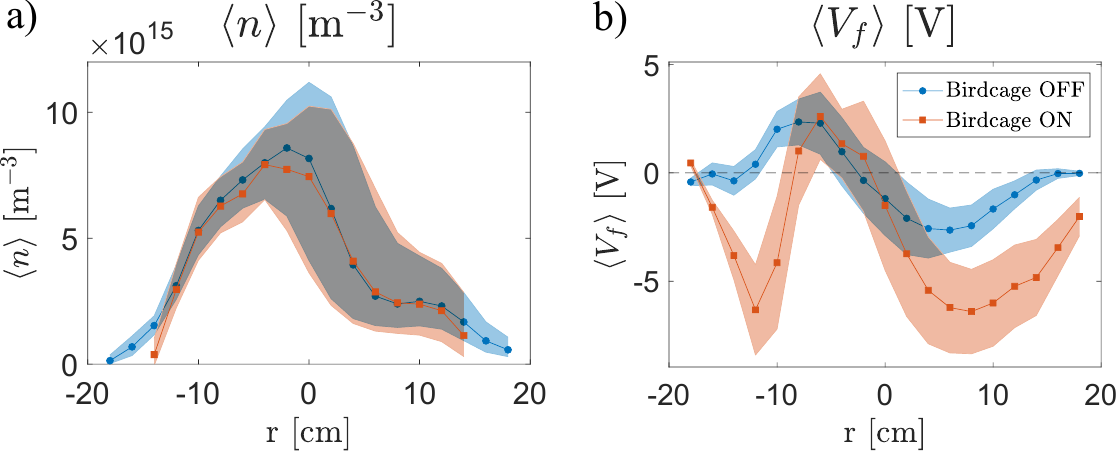}
    \caption{Radial profiles of the density (a) and floating potential (b) measured by Langmuir probe at $z=0$~cm, for $P_{MAG} = 600$~W. The blue curve are measurements with the birdcage antenna off, the red curves with $P_{BC} = 200$~W.}
    \label{langmuir_r_profiles}
\end{figure}

Before looking at the excited helicon waves, \CORR{we look} at the impact of the BCA on the background plasma parameters. Radial profiles of the density $n_e$ and floating potential $V_f$ measured with a Langmuir probe are shown in Fig.~\ref{langmuir_r_profiles} (a) and (b) respectively, both with the BCA \CORR{powered} off (blue curves) and on (red curves). The error bars in \CORR{shaded regions are} the root-mean-square of the measured fluctuations.
The bulk of the magnetron generated plasma is \CORR{located in the region} $-12 \leq r \leq 6$~cm, with $n_e \approx 8 \times 10^{16}$~m$^{-3}$ around $r=-2$~cm. This density profile is not modified by the BCA powered at 200~W. The floating potential however, \CORR{whose absolute value remains} low across the profile with $|V_f| \lesssim 2$~V with the magnetron plasma, is lowered around $r \approx \pm 10$~cm when the birdcage is on, reaching  $V_f \approx -6$~V. Note that this \CORR{happens} close to the glass tube boundary, at $r = \pm 15$~cm, and might indicate an electron temperature increase near the BCA. 
The effect of the BCA \CORR{upon} $n_e$ and $V_f$ depending on the control parameters are discussed in more \CORR{detail} in Sec.~\ref{sec::plasma_birdcage}.

\begin{figure}
    \centering
    \includegraphics[width = 0.98\columnwidth, trim={0in 0in 0in 0in},clip]{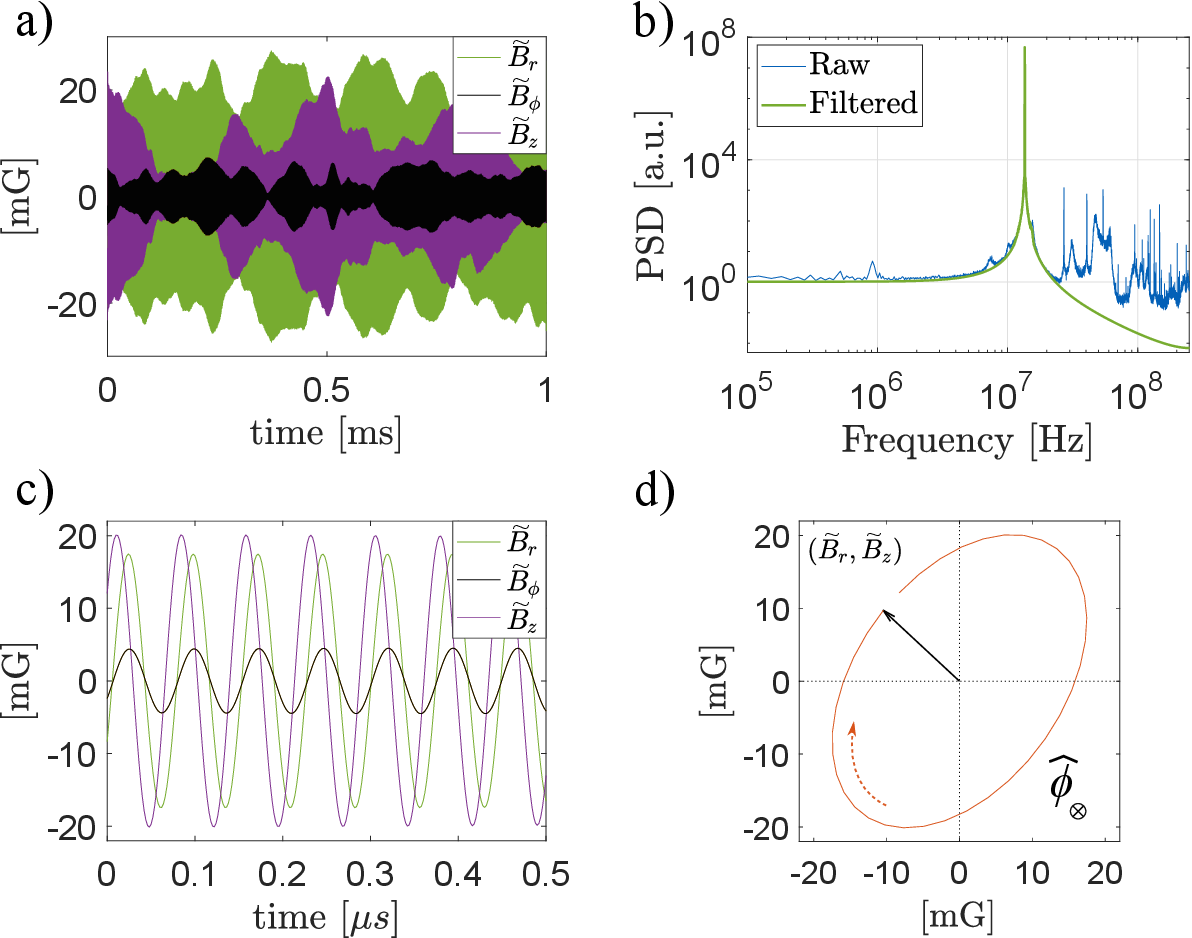}
    \caption{a, c) Time evolution of ($\tilde{B}_r, \tilde{B}_{\phi}, \tilde{B}_z$) measured by B-dot probe at $(r, \phi, z) = (0,  +100 \, ^{\circ}, 0)$. b) Raw and filtered spectra of $\tilde{B}_r$. d) Time evolution of ($\tilde{B}_r$, $\tilde{B}_z$) over one period, showing the polarization of the helicon wave.}
    \label{B_fluct_example}
\end{figure}

Fluctuations of the magnetic field in the 3 directions ($\tilde{B}_r, \tilde{B}_{\phi}, \tilde{B}_z$) are measured with the B-dot probe. An example is shown in Fig.~\ref{B_fluct_example}, \CORR{where each} measurement is acquired over 1~ms, at a sample frequency of 500~MHz. The raw signals are filtered around $f_0 = 13.56$~MHz (Fig.~\ref{B_fluct_example} (b)), and converted to mG (Fig.~\ref{B_fluct_example} (a)) with the calibration procedure detailed in appendix ~\ref{appendix::Bdot}. With the magnetron alone, the measured fluctuations of $B$ are below $\sim 0.3$~mG, which can be considered as the noise level of the B-dot measurements. When the BCA is powered, $B$ fluctuations of the order 10 mG are measured on both sides of the BCA, indicating that the BCA is indeed able to excite and launch \CORR{electromagnetic waves} in the plasma. 
In the magnetic configurations studied here, large-amplitude low frequency fluctuations are observed (see example of Fig.~\ref{B_fluct_example} (a)). The time evolution of the helicon waves is however very stable on the time scale of a few tens of 13.56~MHz periods (Fig.~\ref{B_fluct_example} (c)).

The local polarization as well as the poloidal shape of the helicon wave is \CORR{obtained} by looking at the temporal evolution of ($\tilde{B}_r$, $\tilde{B}_z$), as shown in (Fig.~\ref{B_fluct_example} (d)). \CORR{We observe an elliptical polarization of the waves, which is a characteristic of helicon waves.}
\CORR{In argon and with an external magnetic field of $\approx 780$~G we have $\omega_{ci} \approx 10^5$~Hz and $\omega_{ce} \approx 10^{10}$~Hz. In hydrogen $\omega_{ci} \approx 7\times 10^6$~Hz. We can therefore \CORR{conclude} that the measured waves propagating in TORPEX at a frequency of 13.56~MHz ($\omega = 8.52 \times 10^7$~Hz), that fall in the range $\omega_{ci} \ll \omega \ll \omega_{ce}$, are indeed helicon waves.}

\begin{figure}
    \centering
    \includegraphics[width = 0.98\columnwidth, trim={0in 0in 0in 0in},clip]{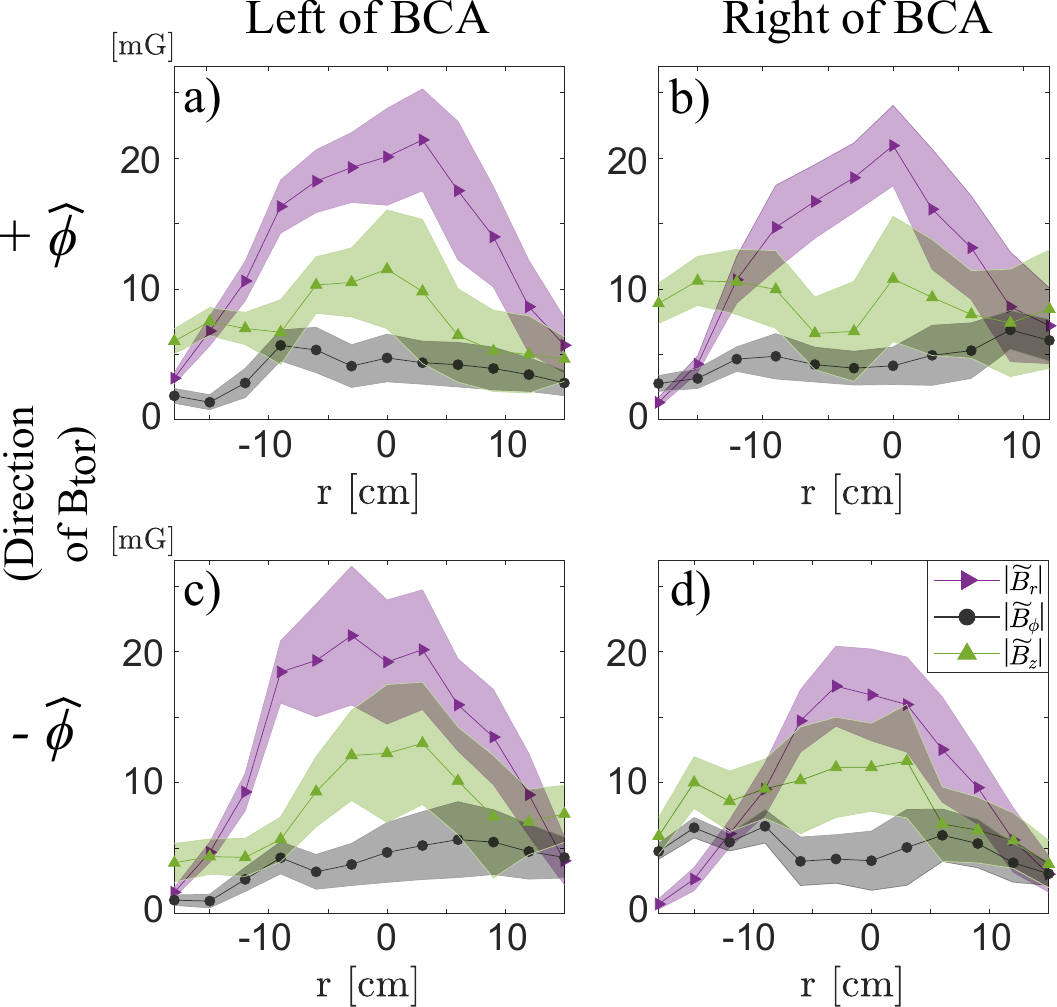}
    \caption{Amplitude of the magnetic field fluctuations measured by the B-dot probe, as a function of the radius, at $z=0$~cm. The measurements are performed at the left (a,c) \CORR{or} right (b,d) of the antenna, as seen from the top, and with the toroidal field $B_{tor}$ along $+\hat{\phi}$ (a,b) \CORR{or} $-\hat{\phi}$ (c,d).}
    \label{B_amp}
\end{figure}

Radial scans are performed with the B-dot probe along $r \in [-18; 18]$~cm by steps of 3~cm, at $z=0$~cm, and both on the left and right sides of the BCA at the toroidal locations $\phi \approx -80 \, ^{\circ}$ and $\phi \approx +100 \, ^{\circ}$, corresponding respectively to the toroidal distances of $\approx -1.4$~m (left) and $\approx +1.8$~m (right) away from the BCA center. 
The amplitudes of the magnetic field fluctuation components ($| \tilde{B}_r |, | \tilde{B}_{\phi} |, | \tilde{B}_z | $) are then computed as the average amplitude over 10 periods. \CORR{The time evolution of these averaged amplitudes} is shown in Fig.~\ref{B_amp}, with the standard deviation of their fluctuations displayed as \CORR{shaded areas}.
Figure~\ref{B_amp} (a) and (b) \CORR{are measurements performed}
 respectively left and right \CORR{of} the BCA.
The radial component $\tilde{B}_r$ is the strongest overall, which is expected as it corresponds to the direction of the magnetic field excited by the BCA (see Fig.~\ref{comsol_sketch} (b)). The shape of $|\tilde{B}_r|$ is centred and almost symmetric around $r=0$~cm; interestingly this does not follow the amplitude of the density, that is off-centred around $r \sim -4$~cm as seen in Fig.~\ref{langmuir_r_profiles} (a).
Note also an increase from the high field side to the low field side of both the radial and toroidal components $\tilde{B}_r$ and $\tilde{B}_{\phi}$, that is likely to be due to the toroidicity of the wave propagation. This is still to be \CORR{further} investigated in future studies.

\begin{figure*}[t]
    \centering
    \includegraphics[width = 1.98\columnwidth, trim={0in 0in 0in 0in},clip]{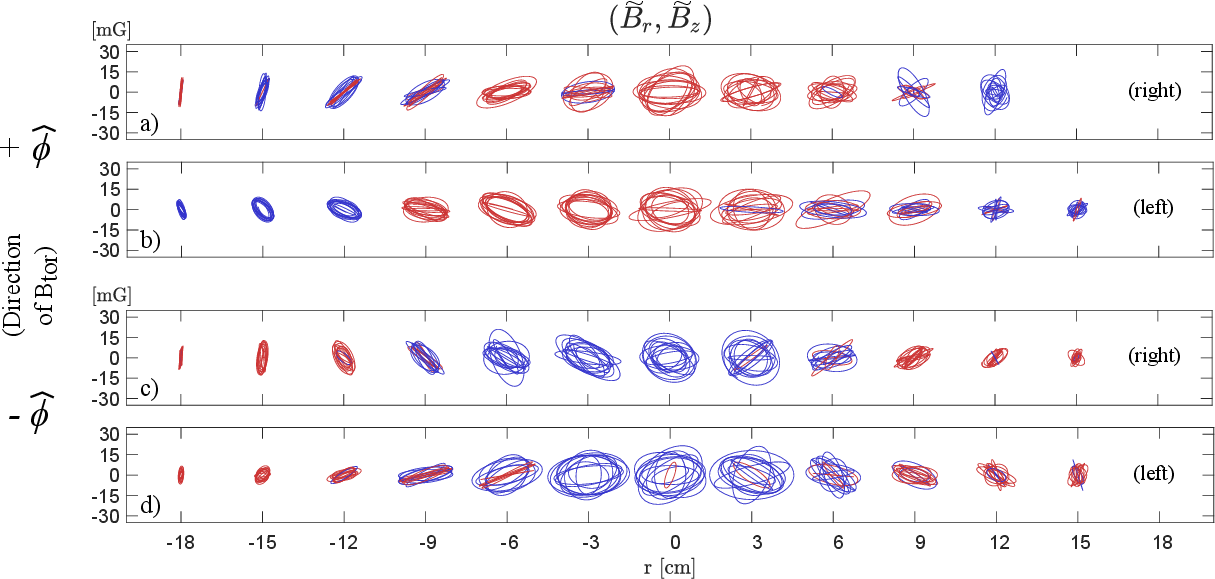}
    \caption{Poloidal components of helicon wave measured with the B-dot along $r$, at $z=0$~cm, on the right (a, c) and left (b,c) of the BCA, for $B_{tor}$ along $\hat{\phi}$ (a,b) or along $- \hat{\phi}$ (c,d). Red corresponds to a positive polarization of the wave with respect to the toroidal axis $\hat{\phi}$ (see Fig~.\ref{B_fluct_example} (d)), blue to a negative one. Ten times are shown. See text for details.}
    \label{B_ellispes}
\end{figure*}

Figures~\ref{B_ellispes} (a) and (b) show the time evolution of ($\tilde{B}_r$, $\tilde{B}_z$),
respectively left and right from the BCA, with 10 different times equally spaced over the 1~ms measurements. Note that for visualization purposes, here the polarization is given with respect to the toroidal axis: red and blue respectively indicate a positive and negative polarization with respect to $\hat{\phi}$.
The local polarization of the helicon wave is positive in the central plasma region and negative at the edges. This global pattern of polarization points to a helicon mode $m = +1$, as indicated by the simple modelling of helicon modes in a uniform cylindrical plasma~\cite{Chen_1991}. Note however that since the latter modeling does not take into account density gradients nor toroidicity, \CORR{which} might have important effects on the wave, a more accurate modelling and comparison to experimental data is still required to identify the helicon modes launched in TORPEX with confidence. \CORR{We note, finally,} that the polarization pattern is almost identical on both sides of the BCA (Fig.~\ref{B_ellispes} (a) and (b)), indicating that the helicon wave launched by the BCA is able to propagate \CORR{throughout} the whole \CORR{toroidal vessel}.

\CORR{To check the robustness of these first measurements,} the direction of the toroidal field is reversed, with $\vec{B}_{tor} = - B_{tor} \hat{\phi}$. The radial profiles of the mean amplitude of ($| \tilde{B}_r |, | \tilde{B}_{\phi} |, | \tilde{B}_z | $) measured at the left and right of the BCA are shown Fig.~\ref{B_ellispes} (c) and (d) respectively. As expected the levels of mean amplitude and fluctuations of $|\tilde{B}|$, as well as the \CORR{corresponding} levels of mean amplitude between the components of $|\tilde{B}|$, are similar to those measured with $\vec{B}_{tor} = B_{tor} \hat{\phi}$ (Fig.~\ref{B_ellispes} (a), (b)).

The magnetic configurations with $\vec{B}_{tor}$ along $\hat{\phi}$ and $-\hat{\phi}$ are symmetric with respect to the toroidal direction.
One could therefore expect a strong \CORR{similarity} between, on the one hand, the profiles measured on the right of the BCA with $\vec{B}_{tor}$ along $\hat{\phi}$ (Fig.~\ref{B_amp} (b)) and the profiles measured on the left of the BCA with $\vec{B}_{tor}$ along $-\hat{\phi}$ (Fig.~\ref{B_amp} (c)) and, on the other hand, between the profiles in Fig.~\ref{B_amp} (a) and (d). But since the BCA input frequency of 13.56~MHz falls by $\sim 1$~MHz away from the BCA resonance, the BCA electromagnetic excitation might not be perfectly symmetric with respect to the axis $\hat{r}$. This is likely to explain the slight discrepancy observed between the pairs of profiles mentioned above.

As for the polarization of the helicon wave, it reverses with the direction of $\vec{B}_{tor}$ as shown in Fig~\ref{B_ellispes} (c) and (d). \CORR{The mode polarization, with respect to the external magnetic field, is therefore identical in both cases $\vec{B}_{tor} = + B_{tor} \hat{\phi}$ and $\vec{B}_{tor} = - B_{tor} \hat{\phi}$.} This gives us further confidence in the ability of the BCA to launch and sustain a $m = +1$ helicon mode in the whole torus of TORPEX.

\section{Plasma generation by the birdcage antenna}
\label{sec::plasma_birdcage}

\CORR{We also explore the ability of the BCA to generate plasma with and without a magnetron-generated background plasma.} The magnetron power $P_{MAG}$ and BCA power $P_{BC}$ are varied in $\{0, 300, 600, 1000\}$~W. Hydrogen and argon SMT plasmas are \CORR{explored}, in the range of pressures $p_0 \in \{0.1, 0.5, 1\} \times 10^{-3}$~mbar.
The density is measured \CORR{using a} Langmuir probe at a position ($r = - 8$~cm, $z = 0$~cm) where the magnetron plasma density is known to be the highest \CORR{from previous studies (not shown)}, and at the toroidal location opposite to the BCA center  ($\phi \approx + 180 \, ^{\circ}$) i.e. where the BCA generated plasma is expected to be weakest.

\begin{figure}
    \centering
    \includegraphics[width = 0.98\columnwidth, trim={0in 0in 0in 0in},clip]{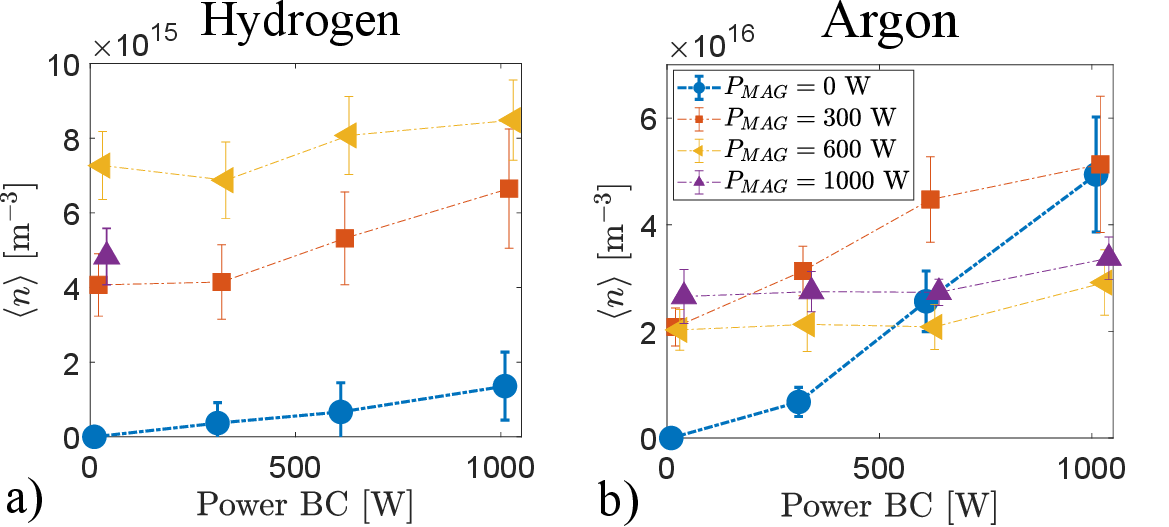}
    \caption{Plasma density measured \CORR{with} Langmuir probe at ($r = - 8$~cm, $z = 0$~cm, $\phi \sim \pi$), for  $P_{MAG}$ and $P_{BC} \in \{0, 300, 600, 1000 \}$~W, at a pressure $p_0 \sim 0.1 \times 10^{-3}$~mbar, in hydrogen (a) and argon (b).}
    \label{langmuir_map}
\end{figure}

Figures~\ref{langmuir_map} (a) and (b) show results for hydrogen and argon, respectively, at a pressure of $0.1 \times 10^{-3}$~mbar. (We note that with hydrogen and a magnetron power of $P_{MAG}=1$~kW, switching on the BCA RF fields causes disturbances to the turbopumps, which prevented the corresponding plasma density measurements. This issue is expected to be solved in the near future by a Faraday shielding of the turbopump entries.)
When the magnetron is on, the plasma density reaches $n_e \approx 4 - 8 \times 10^{15}$~m$^{-3}$ in hydrogen, and $n_e \approx 2 - 5 \times 10^{16}$~m$^{-3}$ in argon, as shown in Fig.~\ref{langmuir_map} for $P_{MAG} = [300; 600; 1000]$~W with the red, yellow and purple curves respectively.
The BCA has a moderate impact on the magnetron-generated plasma with $P_{MAG} = 600$~W and $P_{MAG} = 1000$~W. Indeed with an equal amount of power fed to the BCA than delivered by the magnetron ($P_{BC} = 600$~W and $P_{BC} = 1000$~W respectively), the density increase stays lower than $\sim 25$~\%.
However at a lower magnetron power of $P_{MAG} = 300$~W and in argon, the plasma density increase from the BCA already reaches $\sim  50 $~\% at $P_{BC} = 300$~W, and up to $\sim  150 $~\% at $P_{BC} = 1000$~W.
Finally with the magnetron switched off, the BCA is still able to generate substantial plasma density with respect to the magnetron-generated levels, with $n_e \sim 1.5 \times 10^{15}$~m$^{-3}$ in hydrogen and $n_e \sim 5 \times 10^{16}$~m$^{-3}$ in argon, at $P_{BC} = 1000$~W. \CORR{Note that the BCA is more efficient in argon, which is expected considering in hydrogen the power that is lost into vibrational and rotational energy level excitation, and which does not contribute to ionization.}

\begin{figure}
    \centering
    \includegraphics[width = 0.9\columnwidth, trim={0in 0in 0in 0in},clip]{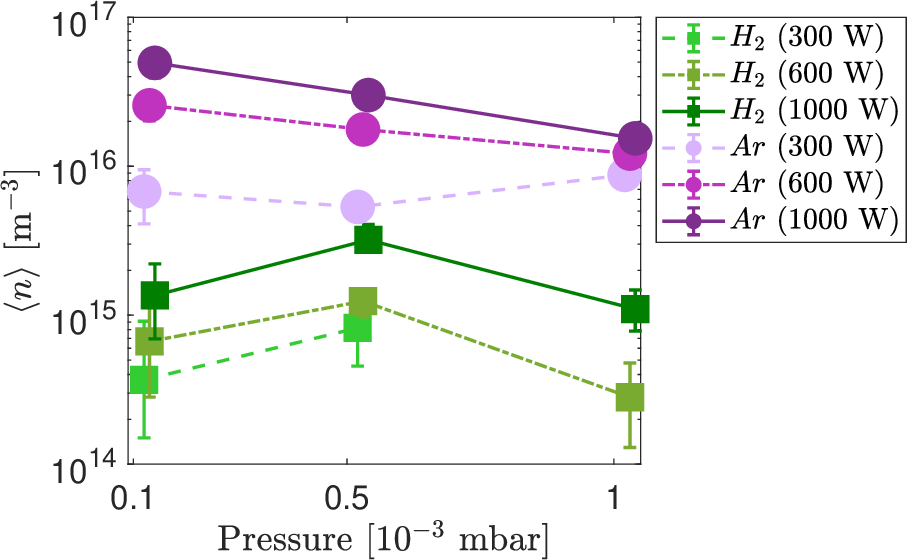}
    \caption{Plasma density measured with Langmuir probe \CORR{at ($r = - 8$~cm, $z = 0$~cm, $\phi \sim \pi$)} as a function of pressure, for hydrogen and argon, \CORR{without magnetron} and for $P_{BC} \in \{300, 600, 1000\}$~W.}
    \label{pressure_scan}
\end{figure}

\CORR{Plasma density} measurements with the BCA alone and with varying pressure are presented in Fig.~\ref{pressure_scan}. The plasma density increases with the BCA power $P_{BC}$, and with argon compared to hydrogen, which extends the observation in Fig.~\ref{langmuir_map} to all pressures $p_0 \in [0.1; 5; 1] \times 10^{-3}$~mbar. \CORR{An increase of pressure is expected to cause a stronger damping of the helicon waves. Since these density measurements are done opposite to the BCA in the chamber, i.e. at a distance $\sim 3$~m from the BCA, we could expect at this location a trend of plasma density decrease with pressure.} Interestingly this trend is only clearly observed in Argon, and for $P_{BC} \geq 600$~W. Globally, the plasma density does not vary significantly when the pressure is increased, staying within a factor $2-5$ from the plasma densities at $p_0 = 0.1 \times 10^{-3}$~mbar.

Note that these power and pressure scans were performed with the Langmuir probe at a single position\CORR{. Therefore} these measurements \CORR{only} provide a first broad overview of the \CORR{ability of the BCA} to generate plasma with or without \CORR{the additional contribution of magnetron-generated microwaves}. \CORR{Since the plasma density and potential shapes vary in 3D when the BCA is used,} a deeper understanding of the impact of the BCA on plasma ionization in TORPEX needs spatially resolved measurements, with radial profiles such as the ones shown in Fig.~\ref{langmuir_r_profiles} or even 2D measurements in ($r,z$), at various toroidal locations, that will be the subject of future work.

\section{Conclusions and outlook}
\label{sec::conclusion}

A 32.5~cm diameter helicon antenna has been designed, \CORR{installed and successfully commissioned} on the toroidal device TORPEX. An alternative design to the most commonly used half-helical antenna design was chosen, consisting in a resonant network of inductive legs connected in parallel by capacitors. This so-called birdcage antenna has the advantage of \CORR{operating with} low input voltage and current, making it easily scalable to larger dimensions than the $\sim 10$~cm diameter sources \CORR{typically used in} low temperature plasma \CORR{linear} devices. The sinusoidal distribution of currents in the legs at resonance excites transverse $m = 1$ magnetic field, making the BCA an efficient helicon source. The large surface coverage \CORR{of the legs} around the dielectric tube also makes this antenna \CORR{efficient at generating} and inductively coupling to the plasma, even at \CORR{a 
(low)} pressure of $0.1 \times 10^{-3}$~mbar both in argon and in hydrogen. Finally, \CORR{owing to an input inductance $Z$ with }$\text{Re}(Z) \sim 100$~$\Omega$ and $|\text{Im}(Z)| \lesssim 50$~$\Omega$ \CORR{close to} the resonance and in the presence of plasma, the matching of the BCA is straightforward and was \CORR{easily} achieved with a type T matching-box.

\CORR{We stress that for all the measurements presented in sections~\ref{sec::propag_helicon} and \ref{sec::plasma_birdcage}, no adjustment to the BCA was needed. Once built and \CORR{installed} in TORPEX the BCA could be used without any further modification or RF coupling issues. This simplicity of use (once passed the design and building phases) \CORR{summarizes the essential benefits} brought by the BCA on a large scale set-up such as TORPEX.}

Helicon waves \CORR{are} launched and sustained \CORR{around the entire toroidal vessel} of TORPEX, as \CORR{demonstrated} by B-dot measurements on both sides of the BCA. The wave polarization pattern \CORR{suggests} that a mode $m=+1$ is dominant, by comparing \CORR{our measurements} to the simple modelling of helicon modes in a uniform cylindrical plasma. Further measurements and comparisons with models are required to confirm this mode identification. \CORR{These} will include measurements of the 2D profiles of the helicon wave amplitude across the poloidal cross section of TORPEX, as well as measurements of the toroidal wave number.

\CORR{More generally, } thanks to its numerous available diagnostics~\cite{Theiler_2011, Baquero_2016, thesis_Manke_2020}, wide set of coils, and newly installed BCA, TORPEX is now a flexible and insightful test bed for the experimental study of the fundamental properties of helicon waves in toroidal geometry. 

At a low power of 200~W, the BCA \CORR{is} shown to be able to launch helicon waves in a pre-existing argon plasma of density $n_e \sim  10^{15}-10^{16}$~m$^{-3}$, without \CORR{leading to modifications of} the plasma density profile. This is an interesting feature for future studies in TORPEX, \CORR{in which we plan to investigate} the impact of \CORR{resonant perturbations from} helicon waves on the propagation of fast ions in a background turbulent plasma.

At powers up to 1~kW, plasma density measurements demonstrate the ability of the BCA to substantially modify the plasma equilibrium, as well as to generate\CORR{, without contributions from the magnetron,} a plasma density up to $5 \times 10^{16}$~m$^{-3}$.

The \CORR{constraint imposed by the magnetron on the allowed values of magnetic fields (see Sec.~\ref{sec::intro})} is therefore removed.
This opens the possibility to extend former studies carried out in TORPEX, such as blob transport mechanisms, with the added perspective of a varying magnetic field. 
The BCA moreover opens the way to entirely new experimental studies on TORPEX, such as transition to turbulence, or the impact of the magnetic field to the transport properties of fast ions.
Comparing experiments with simulations as a mean for code validation will also now be possible in regimes previously inaccessible in TORPEX.

\section*{Acknowledgements}

This work has been carried out within the framework of the EUROfusion
Consortium, via the Euratom Research and Training Programme (Grant
Agreement No 101052200 — EUROfusion) and funded by the Swiss State
Secretariat for Education, Research and Innovation (SERI). Views and
opinions expressed are however those of the author(s) only and do not
necessarily reflect those of the European Union, the European
Commission, or SERI. Neither the European Union nor the European
Commission nor SERI can be held responsible for them. This work was also supported in part by the Swiss National Science Foundation.

\appendix 

\section{B-dot calibration}
\label{appendix::Bdot}

Let us consider magnetic field fluctuations ($B_x(t)$, $B_y(t)$, $B_z(t)$) at the loaction of the B-dot probe. The magnetic field $B_x(t)$ mainly induces a \CORR{voltage} at the terminals of the B-dot X-coil; this signal is denoted $\varepsilon_{XX}(t)$. However, since the \CORR{coil orientation is not perfect}, the measured signal at the X coil terminal \CORR{might} also be due to induced \CORR{voltages} from the $B_y(t)$ and $B_z(t)$ components, \CORR{voltages} respectively denoted $\varepsilon_{XY}(t)$ and $\varepsilon_{XZ}(t)$. We can write:
\begin{align*}
    V_x(t) & = \varepsilon_{XX}(t) + \varepsilon_{XY}(t) + \varepsilon_{XZ}(t).
\end{align*}

A similar statement holds for measured signals $V_y(t)$ and $V_z(t)$. Then each \CORR{voltage} $\varepsilon_{ij}(t)$, induced in coil $i$ by the magnetic field component $j$ can be written as $\varepsilon_{ij}(t) = f_{ij}(B_j(t))$,
with $f_{ij}$ a function changing the amplitude and phase of each  $B_j(t)$ frequency components, respectively by a factor $a_{ij}(\omega)$ and a phase $\phi_{ij}(\omega)$.
This results in the following set of equations :
\begin{align}
    \begin{cases}
    V_x(t)  & = f_{XX}(B_x(t)) + f_{XY}(B_y(t)) + f_{XZ}(B_z(t)) \\
    V_y(t)  & = f_{YX}(B_x(t)) + f_{YY}(B_y(t)) + f_{YZ}(B_z(t)) \\
    V_z(t)  & = f_{ZX}(B_x(t)) + f_{ZY}(B_y(t)) + f_{ZZ}(B_z(t))
    \end{cases}
    \label{eq::3x3_temporal}
\end{align}

\CORR{Going to} the spectral domain, and \CORR{looking} only at the components along a given frequency $\omega_0$, Eq.~\eqref{eq::3x3_temporal} become
\begin{align*}
    \begin{pmatrix}
    \hat{V}_x
    \\
    \hat{V}_y
    \\
    \hat{V}_z
    \end{pmatrix}
    =  \underbrace{\begin{pmatrix}
    a_{XX} e^{i \phi_{XX}}  &
    a_{XY} e^{i \phi_{YX}}  &
    a_{XZ} e^{i \phi_{ZX}} 
    \\
    a_{YX} e^{i \phi_{XY}}  &
    a_{YY} e^{i \phi_{YY}}  &
    a_{YZ} e^{i \phi_{ZY}} 
    \\
    a_{ZX} e^{i \phi_{XZ}}  &
    a_{ZY} e^{i \phi_{YZ}}  &
    a_{ZZ} e^{i \phi_{ZZ}}
    \end{pmatrix}}_{M}
    \begin{pmatrix}
    \hat{B}_x
    \\
    \hat{B}_y
    \\
    \hat{B}_z
    \end{pmatrix}.
\end{align*}

To determine the coefficients of the matrix $M$, the B-dot signal is measured for magnetic fields aligned along one direction $x$, $y$, $z$. For instance, along $x$,
\begin{align}
    \begin{pmatrix}
    \hat{V}_{xx}^0
    \\
    \hat{V}_{xy}^0
    \\
    \hat{V}_{xz}^0
    \end{pmatrix}
    & = \begin{pmatrix}
    a_{XX} e^{i \phi_{XX}}  &
    a_{XY} e^{i \phi_{YX}}  &
    a_{XZ} e^{i \phi_{ZX}} 
    \\
    a_{YX} e^{i \phi_{XY}}  &
    a_{YY} e^{i \phi_{YY}}  &
    a_{YZ} e^{i \phi_{ZY}} 
    \\
    a_{ZX} e^{i \phi_{XZ}}  &
    a_{ZY} e^{i \phi_{YZ}}  &
    a_{ZZ} e^{i \phi_{ZZ}}
    \end{pmatrix}
    \begin{pmatrix}
    \hat{B}_0
    \\
    0
    \\
    0
    \end{pmatrix} \label{eq::calibration_B0_1} \\
    & = \begin{pmatrix}
    a_{XX} e^{i \phi_{XX}} \hat{B}_0 \\
    a_{XY} e^{i \phi_{YX}} \hat{B}_0  \\
    a_{XZ} e^{i \phi_{ZX}} \hat{B}_0 
    \end{pmatrix}.
\end{align}

\noindent Then the signal measured by any magnetic field is given by
\begin{align*}
    \begin{pmatrix}
    \hat{V}_x
    \\
    \hat{V}_y
    \\
    \hat{V}_z
    \end{pmatrix}
    = \frac{1}{|\hat{B}_0|}
    \begin{pmatrix}
    V_{XX}^0 e^{i \phi_{XX}}  &
    V_{XY}^0 e^{i \phi_{YX}}  &
    V_{XZ}^0 e^{i \phi_{ZX}} 
    \\
    V_{YX}^0 e^{i \phi_{XY}}  &
    V_{YY}^0 e^{i \phi_{YY}}  &
    V_{YZ}^0 e^{i \phi_{ZY}} 
    \\
    V_{ZX}^0 e^{i \phi_{XZ}}  &
    V_{ZY}^0 e^{i \phi_{YZ}}  &
    V_{ZZ}^0 e^{i \phi_{ZZ}}
    \end{pmatrix}
    \begin{pmatrix}
    \hat{B}_x
    \\
    \hat{B}_y
    \\
    \hat{B}_z
    \end{pmatrix} .
    \label{}
\end{align*}

\CORR{From this expression, we conclude} :

\begin{align*}
    \begin{pmatrix}
    \hat{B}_x
    \\
    \hat{B}_y
    \\
    \hat{B}_z
    \end{pmatrix}
    =
    |\hat{B}_0| \cdot A^{-1} \cdot
    \begin{pmatrix}
    \hat{V}_x
    \\
    \hat{V}_y
    \\
    \hat{V}_z
    \end{pmatrix}.
    \label{}
\end{align*}

The calibration measurements illustrated by Eq.~\eqref{eq::calibration_B0_1} are performed using a 10~cm diameter dedicated birdcage antenna. After a B-dot measurement of ($V_x, V_y, V_z$), each raw signal $V_i$ is filtered around $f_0 = 13.56 \pm 1$~MHz using Matlab bandpass filter with a default transition band steepness of 0.85. Note that the spectrum of the filtered signal (green curve in Fig.~\ref{B_fluct_example} (b)) strongly peaks at $f_0$, and \CORR{falls by} over 7 orders of magnitude at $f_0 \pm 0.5$~MHz. Inside this frequency range, the calibration can still be considered valid; away from this frequency range, the frequency amplitude of the filtered signal is negligible. The calibration procedure is therefore not only applied to the 13.56~MHz component, but to the full filtered signals: this allows most of the low frequency fluctuations of the original signals to be kept.

\section{Equation solved in COMSOL}
\label{appendix::COMSOL}

The equations that are used in COMSOL simulations are derived here. We assume cold ions, and consider the electron momentum equation
\begin{align*}
    m_e n_e \partial_t \vec{u}_e & = -e n_e \vec{E} - e n_e \vec{u}_e \times \vec{B} - m_e n_e \nu \vec{u}_e \\
    \frac{m_e}{n} \underbrace{(\nu + i\omega)}_{\gamma} \vec{u}_e & = - \vec{E} - \vec{u}_e \times \vec{B} .
    \label{eq::momentum}
\end{align*}

\noindent Neglecting the \CORR{inertia of the ions}, the electric current can be written like $\vec{j} = -e n_e \vec{u}_e $. \CORR{Using then} $\vec{u}_e = -\frac{\vec{j}}{e n}$ with $\vec{\omega}_c = e \vec{B}/m_e$,  the momentum equation reads
\begin{equation}
    \gamma \vec{j} = \varepsilon_0 \omega_p^2 \vec{E} - \vec{j} \times \vec{\omega}_c .
    \label{eq::momentum-2}
\end{equation}

\noindent \CORR{We now perform the cross product of} $\times \vec{\omega}_c$ to Eq.~\eqref{eq::momentum-2}, then substitute in the left hand side the expression of $\vec{j} \times \vec{\omega}_c$ given by Eq.~\eqref{eq::momentum-2}, and obtain
\begin{align*}
    \gamma^2 \vec{j}  - \gamma \varepsilon_0 \omega_p^2 \vec{E} & = \varepsilon_0 \omega_p^2 \vec{E} \times \vec{\omega}_c - (\vec{j} \times \vec{\omega}_c) \times \vec{\omega}_c .
\end{align*}

\noindent By noting that $(\vec{j} \times \vec{\omega}_c) \times \vec{\omega}_c = - \omega_c^2 \vec{j}_{\perp}$ (the direction $\parallel$ and $\perp$ being defined with respect to $\vec{B}$), this last expression can be rewritten like
\begin{align*}
    \gamma^2 \vec{j}  + \omega_c^2 \vec{j}_{\perp}  = \varepsilon_0 \omega_p^2 (\gamma \vec{E} + \vec{E} \times \vec{\omega}_c),
\end{align*}

\noindent which, for the simple case $\vec{B} = B \vec{e}_z$, hence with $\vec{\omega}_c = \omega_c \vec{e}_z$, yields
\begin{align*}
    \Bigg( \Bigg.\begin{matrix}
    (\gamma^2 + \vec{\omega}_c^2) j_x \\
    (\gamma^2 + \vec{\omega}_c^2) j_y \\
    \gamma^2 j_z
    \end{matrix} \Bigg. \Bigg)
    & =  \varepsilon_0 \omega_p^2
    \Bigg( \Bigg.\begin{matrix}
    \gamma & \omega_c & 0 \\
    -\omega_c & \gamma & 0 \\
    0 & 0 & \gamma
    \end{matrix}\Bigg. \Bigg)
    \Bigg( \Bigg.\begin{matrix}
    E_x \\
    E_y \\
    E_z
    \end{matrix}\Bigg. \Bigg)
\end{align*}

\noindent This result can be written like
\begin{align*}
    \vec{j} = \boldsymbol{\sigma} \cdot \vec{E}.
\end{align*}

\noindent with $\boldsymbol{\sigma}$ as defined in subsection~\ref{subsec::num_simu}. \CORR{Combining this result with Maxwell's equations}
then yields the final equation for $\vec{E}$ provided in subsection~\ref{subsec::num_simu}.

\bibliography{mybib_birdcage_torpex}{}
\bibliographystyle{unsrt}

\end{document}